\documentclass[12pt]{article}

\usepackage{graphicx}
\usepackage{amssymb,amsfonts,amsmath}
\usepackage{epstopdf}
\usepackage{hyperref}
\usepackage{latexsym}
\usepackage{marvosym}
\usepackage{indentfirst}
\usepackage{bbm}

\begin{document}

\begin{titlepage}
\bigskip
\rightline{}
\bigskip\bigskip\bigskip\bigskip
\centerline{\LARGE   {  De Sitter Entropy from a  Lower }}
\vspace{5pt}
\centerline{\LARGE   {   Dimensional Black Hole}}
\bigskip\bigskip
\bigskip\bigskip
 \centerline{\large Cesar Arias,${}^1$ Rodrigo Aros,${}^2$
   and Nelson Zamorano${}^3$}
\vspace{5pt}
       \center{
${}^{1,2}${\em \normalsize
Departamento de Ciencias F\'isicas\\
 Facultad de Ciencias Exactas \\
Universidad Andres Bello\\
Santiago de Chile.}\\
\vspace{10pt}
${}^{3}${\em \normalsize
Departamento de F\'isica\\
 Facultad de Ciencias F\'isicas y Matem\'aticas \\
Universidad de Chile\\
Santiago de Chile.}\\
}

\vspace{15pt}
\begin{abstract}
\vspace{8pt}
\baselineskip=15pt
An alternative method to compute the entropy of de Sitter spacetime for dimensions higher than or equal to five is presented.
It is shown that de Sitter entropy may be obtained from the Bekenstein-Hawking entropy  of a lower-dimensional black hole. 
This result follows from 
the existence of a one-to-one correspondence between de Sitter and a spacetime containing a black hole localized on a  $p$-brane. Under this correspondence and due to the holographic principle, the entropy of a five-dimensional de Sitter space may be obtained from that of a BTZ black hole localized on a 2-brane. For dimensions higher than five, de Sitter entropy is identified with the entropy of a Schwarzschild-de Sitter black hole localized on a ($d-$3)-brane.
\end{abstract}

\vspace*{2cm}
\begin{flushleft}
~\\
\hspace{27pt}{${}^{1}$\href{mailto:ce.arias@uandresbello.edu}{\footnotesize  ce.arias@uandresbello.edu}}\\
\hspace{27pt}{${}^{2}$\href{mailto:raros@unab.cl}{\footnotesize raros@unab.cl}}\\
\hspace{27pt}{${}^{3}$\href{mailto:nzamora@ing.uchile.cl}{\footnotesize nzamora@ing.uchile.cl}}\\
\end{flushleft}

\end{titlepage}

\baselineskip=18pt

\setcounter{equation}{0}

\newpage
\tableofcontents
\vspace{.4cm}
\bigskip\hrule

\section{Introduction}
The latest observational evidence indicates that the Universe we live in is expanding at an accelerated rate \cite{Supernovae1, Supernovae2}. This means that our cosmos might currently be described by a de Sitter geometry. A $d$-dimensional de Sitter spacetime ($dS_d$) is the maximally symmetric Lorentzian Einstein manifold with positive curvature, characterized by a positive cosmological constant $\Lambda > 0$. Because it describes an accelerated Universe, an observer in de Sitter space does not have access to the whole space and, indeed, can only see part of it. This gives rise to a cosmological horizon: any light ray beyond  it can never reach to the observer.  Consistently, the geometry of $dS_d$ exhibits compact spacelike slices, which implies two important facts. Unlike its anti-de Sitter counterpart, the notion of spatial infinity becomes ill defined on $dS_d$. Moreover, the physical states of any consistent quantum theory of gravity on $dS_d$ must be defined on a finite-dimensional Hilbert space \cite{Banks, Witten1}.

In 1977, Gibbons and Hawking \cite{GibbonsHawking} showed that the $dS_d$ cosmological horizon displays the same thermodynamical features as the event horizon of a black hole. Remarkably, its entropy matches exactly the Bekenstein-Hawking prescription \cite{Bekenstein, Hawking}.  Despite this similitude and unlike the black hole case, the $dS_d$  horizon is observer-dependent and therefore  is not clear what are the fundamental properties shared by both types of horizons. It is not known why the two entropies are exactly one-quarter of the area (in Planck units) of the respective horizon.

Despite the lack of a universal prescription, it has been possible to compute from first principles the entropy of some specific black holes \cite{Carlip, StromingerVafa, Strominger}. However, some problems arise when trying to applied similar procedures to $dS_d$ space, with the exception of the (2+1)-dimensional case \cite{MaldacenaStrominger}. One of these problems concern the asymptotic structure of the space. There is no notion of spatial infinity within $dS_d$ so that the asymptotia must be in the past and the future. Another obstacle is that $dS_d$ does not admit a supersymmetric extension.  It is not possible to extend the isometries of de Sitter space into a supergroup having a unitary representation \cite{Peter}. Furthermore, even at the classical level, there is no supergravity compactifications leading to de Sitter space of any dimension. That is, $dS_d \times \mathcal M$, with $\mathcal M$ some arbitrary compactification manifold, is not a background  for ten-, nor eleven-, dimensional supergravity \cite{deWit, nogo}.

In this article we study the problem of the $dS_d$ entropy for dimensions higher than or equal to five. According to the holographic principle by 't Hooft and Susskind  \cite{tHooft, Susskind}, the microscopic degrees of freedom that result in the entropy of $dS_d$ should be localized on its cosmological horizon, $\mathcal H$. This is a $(d-2)$-dimensional surface with the topology of a $(d-2)$-sphere, and it represents the spatial boundary of the $dS_d$ geometry. With this configuration in mind, we will address two questions here. Is there some connection between the entropy of de Sitter space with the entropy of a black hole? If so, can we formulate the entropy problem in such a way that this connection becomes manifest? Looking for an answer to these questions, we will argue that the entropy of $dS_d$ can be related to that of a lower dimensional black hole, through a \emph{topology change}. The basic idea lies in the fact that the $dS_d$ spacetime can be mapped into another geometry, having a different topology, namely, $\widetilde{dS_d}$
\begin{eqnarray}
\mathfrak{F} : dS_d &\longrightarrow&  \widetilde{dS_d}  \nonumber \\
                    \mathcal H  &\longmapsto& \mathfrak{F} (\mathcal H)
\label{intro}                    
\end{eqnarray}
such that, under $\mathfrak{F}$, we can identify the degrees of freedom carried by the $dS_d$ horizon $\mathcal H$ with those carried by its image 
$\mathfrak{F} (\mathcal H) \subseteq \widetilde{dS_d}$. Moreover, the geometry  of $\widetilde{dS_d}$ is that of black hole localized on the  $\mathfrak{F} (\mathcal H)$ hypersurface. Therefore, due to the proposed correspondence and in light of the holographic principle, we show that computing the entropy of de Sitter space in the left side of Eq.(\ref{intro}) is equivalent to computing the Bekenstein-Hawking entropy of a black hole on the $\widetilde{dS_d}$ side.

The paper is organized into three sections.  In section \ref{setup}, the construction of the correspondence (\ref{intro}) is performed.  Starting with the $dS_d$ geometry, a change in the topology is carried out, leading to $\widetilde{dS_d}$. Hence, we will have two different backgrounds,  solving the same equations of motion, and that can be mapped one into the other through a given set of transformations.  In particular, the cosmological horizon of the original $dS_d$ space will be mapped into a $(d-2)$-dimensional timelike surface whose induced metric describes a black hole, $ \mathcal H  \mapsto \mathfrak{F} (\mathcal H)$.  In section \ref{localization} it is shown that the hypersurface $\mathfrak{F} (\mathcal H)$ defines a gravitational source, or a $p$-brane, located at the origin of $\widetilde{dS_d}$. Specifically, a conical defect can be induced at that point, acting as a $\delta$-type  contribution to the energy-momentum tensor of the system. Then, the configuration of the image space can be thought of as a codimension-two brane black hole.  Section (\ref{entropy}) is devoted to  computing the entropy of de Sitter space using the aforementioned construction. We show that the $dS_d$ entropy matches exactly the Bekenstein-Hawking entropy of the black hole localized on $\mathfrak{F} (\mathcal H)$, provided a fixed angular deficit is introduced along the transverse directions to the brane. We conclude in section (\ref{CR}) with some remarks and comments regarding our results, including a description of further developments and open problems. We will adopt units such that  $\hbar=c=1$ but keeping Newton's constant $G\neq1$ throughout.


\section{Correspondence}
\label{setup}
In this section, we proceed to construct the map $\mathfrak{F} : dS_d \rightarrow \widetilde{dS}_{d}$. It will be shown that, starting from de Sitter space, one can implement a set of transformations that change the topology of the original geometry, resulting into a new family of asymptotically de Sitter backgrounds for $d\geq 4$. 

Let us begin our prescription considering the $d$-dimensional de Sitter spacetime in static coordinates:
\begin{equation}
ds^2= - \left( 1- \frac{R^2}{L^2} \right)  dT^2  + \frac{dR^2}{ \left( 1- \frac{R^2}{L^2} \right) }  + R^2 d\Omega^{2}_{d-2}.
\label{dSd}
\end{equation}
Here $d\Omega^{2}_N$ denotes the unitary $N$-sphere line element, and $L$ denotes the $dS_d$ radius, related with the $d$-dimensional cosmological constant
$\Lambda > 0$, through $L^2=\frac{(d-1)(d-2)}{2 \Lambda}$. The above set of coordinates cover only the patch defined by $0\leq R \leq L$, and represents an observer located at $R=0$ surrounded by a cosmological horizon at $R=L$. Each spacelike surface at $R=R_0=$ constant, on the other hand, has the topology of a $(d-2)$-sphere of radius $R_0$.

We define  now the action of $\mathfrak{F}$ on $dS_d$, introducing the following set of transformations in Eq.(\ref{dSd}) : first, a coordinate change $R=R(z)$ such that the origin $R=0$ is mapped into $z \rightarrow \infty$ and the cosmological horizon $L$ is mapped to $z=0$. In this way, the new coordinate will cover the whole interval $0\leq z < \infty$. Explicitly,
\begin{equation}
R= \frac{L}{\cosh (z/L)} .
\label{R(z)}
\end{equation}
Secondly, still in Eq.(\ref{dSd}), we replace the horizon $\mathcal H$, defined by the spacelike hypersurface $R=L$,  by a $(d-2)$-dimensional spacetime, namely,\footnote{About the notation : here and in what follows, the \emph{hat} will denote objects intrinsically $(d-2)$-dimensional, and, therefore, the quantities with this label will depend only on the $x$ coordinates, which ensures the Lorentz invariance of the subspace defined by $\hat{g}$. In this way, Greek indices run from $0$ to $(d-2)-1$, while capital ones will run over the whole space.}
$d\hat{s}^2_{d-2}$
\begin{eqnarray}
L^2 d\Omega^2_{d-2} \rightarrow d\hat s^2 &=& \hat{g}_{\mu\nu}(x) dx^{\mu}dx^{\nu}   \nonumber \\
 								    &=&- f(r) dt^2 + \frac{dr^2}{f(r)} + r^2 d\Omega^2_{d-4}.
\label{brane}
\end{eqnarray}
Note that the above change includes a Wick rotation, and it truly represents a change of topology. Later, depending on the resulting submanifold in the rhs of Eq.(\ref{brane}), this transformation may be written in a clearer way in terms of the isometry groups involved in it 
[c.f. Eq.(\ref{group})].

Finally, to preserve the Lorentzian signature of the whole space, a second Wick rotation is in order:
\begin{equation}
T \rightarrow iL \phi.
\label{Wick}
\end{equation}

Once the sequence of transformations (\ref{R(z)})-(\ref{Wick}) into Eq.(\ref{dSd}) has been implemented, the resulting line element is
\begin{equation}
ds^2 = \frac{1}{\cosh^2(z/L)} \left[ dz^2 + L^2 \sinh^2(z/L) d\phi^2 +  \hat{g}_{\mu\nu}(x) dx^{\mu}dx^{\nu}  \right].
\label{new}
\end{equation}
This metric solves Eintein's equations plus a cosmological constant $G_{MN} + \Lambda g_{MN}=0$, provided that
\begin{equation}
 f(r) = \beta-\frac{\mu}{r^{d-5}} - \frac{r^2}{L^2} =: f_{d}(r).
 \label{f}
 \end{equation}
In the last expression, $\beta$ and $\mu$ are two parameters depending on the dimension of the spacetime. $\beta$ is an arbitrary constant for $d=4,5$, but it must be fixed to the unit for $d\geq 6$. On the other hand, $\mu$ is related with the black hole structure of the $\hat{g}$-subspace. To our best knowledge, the spacetime defined by Eqs.(\ref{new}) and (\ref{f}) represent a new family of asymptotically de Sitter solutions to the Einstein field equations. Its geometry, therefore, remains unknown.

It is important to point out that the transformations displayed by Eqs.(\ref{brane}) and (\ref{Wick}) are not  diffeomorphisms. Indeed, the spaces (\ref{dSd}) and (\ref{new})-(\ref{f}) are topologically inequivalent. This can be readily checked by computing some curvature invariants. For instance, while Ricci scalar and the full contraction of the Ricci tensor are the same for both spaces,
 \begin{equation}
R  =  \frac{d(d-1)}{L^2}  \, \, \, ; \, \, \,   R_{AB}R^{AB} = \frac{d(d-1)^2}{L^4} ,
\label{Riccis}
\end{equation}
the Kretschmann scalars differ. For the line element (\ref{new}),
\begin{eqnarray}
K:=R_{ABCD}R^{ABCD} &=&   \frac{2d(d-1)}{L^4}  + \mu^2 \ (d-3)(d-4)^2(d-5)\,\frac{\cosh^4(z/L)}{r^{2(d-3)}}   \nonumber \\
                                              &=:& K_{dS} + \mu^2\, \xi(d)\, \frac{\cosh^4(z/L)}{r^{2(d-3)}},
\label{K}
\end{eqnarray}
whereas for $dS_d$ space, the last term vanishes. In this case, the new space described by Eqs.(\ref{new}) and (\ref{f}) is not a maximally symmetric,  and, furthermore, for
 $d\geqslant 6$, it has two curvature singularities at $r=0$ and $z \rightarrow \infty$ [the image of the $dS_d$ origin under Eq.(\ref{R(z)})]. 
 The first singularity reveals the existence of a \emph{black hole}.

In what follows, the new geometry (\ref{new})-(\ref{f}) is called image space and denoted by  $\widetilde{dS}_{d}$. The table below summarizes the correspondence between spaces. \\

\begin{center}
\begin{tabular}{| c |c |}
\hline $dS_d$ & $ \widetilde{dS}_{d}$ \\
\hline Coordinates $(R,T, \Omega_{d-2})$ & Coordinates $(z,\phi, x^{\mu})$ \\
\hline Origin $R=0$                                         &     $z \rightarrow  \infty$                \\
\hline Horizon $R=L$                                      &      $z =0$                                        \\
   (spacelike surface $ L^2 d\Omega^{2}_{d-2}$)   &   (timelike surface $d\hat{s}^2$)                         \\
\hline   $K=K_{dS}$    &       $K=K_{dS} + \mu^2\,\xi(d) \,  \frac{\cosh^4(z/L)}{r^{2(d-3)}}   $      \\
\hline
\end{tabular}
\end{center}


\section{Brane Localization}
\label{localization}
In this section, we show that the spacetime $\widetilde{dS}_{d}$ can be seen as a codimension two brane black hole\footnote{For a review of brane black holes, see Refs.\cite{GregoryReview, Kanti} and references in it. Previous literature regarding the codimension-two scenario includes Refs.\cite{RubakovShaposhnikov, GherghettaShaposhnikov, CarrollGuica, CanadaGroup, DurhamGroup, Burgess}. }. More precisely, the
$(d-2)$-dimensional surface $\mathfrak{F} (\mathcal H) \subseteq \widetilde{dS_d}$ defined by $z=0$, can be localized on a conical defect in $d$ dimensions, acting as a gravitational source. The first clue of this configuration, is the asymptotic behavior of the metric (\ref{new}) around the origin
\begin{equation}
ds^2|_{z \rightarrow 0}  \simeq  dz^2 +  z^2 d\phi^2  + \hat{g}_{\mu\nu}(x) dx^{\mu} dx^{\nu}
\end{equation}
so that if we identify the angular coordinate $\phi \rightarrow \alpha \phi$ with $\alpha \neq 1$, a conical singularity will be induced, defined by the deficit angle
\begin{equation}
\Delta \phi = 2\pi (1-\alpha).
\label{deficit}
\end{equation}
This angular deficit can be interpreted as the geometrical effect of a gravitational source  \footnote{ This is the codimension-two analog of considering the discontinuity in the extrinsic curvature, in the codimension-one case,  as a $\delta$-source of energy-momentum at the brane position.}, with an energy-momentum (density) proportional to $\Delta \phi$. The argument to conclude this goes as follows \cite{DurhamGroup}: consider a $d$-dimensional spacetime characterized by the metric ansatz in normal (Gaussian) coordinates,
\begin{equation}
ds^2 =  d\rho^2 + B^2 (\rho) d\phi^2 + W^2(\rho)\: \hat{g}_{\mu\nu}(x) dx^{\mu} dx^{\nu}
\label{normal}
\end{equation}
then, if the function $B=B(\rho)$ admits, for $\rho \rightarrow 0$, the expansion
\begin{equation}
B(\rho)= \alpha \rho + O(\rho^2)
\label{Lexpansion}
\end{equation}
with $\alpha \neq 1$, we have a conical singularity in the $(\rho, \phi)$ plane, located at $\rho=0$. The space in the vicinity of this conical singularity will be regular, that is, any invariant constructed from de Riemann curvature tensor will remain finite around $\rho\rightarrow0$, if we demand the following boundary conditions (primes indicates derivatives with respect to $\rho$):
\begin{eqnarray}
B |_{\rho=0} &=&0 \,  ; \, \,  \, \, B^{\, \prime} |_{\rho=0} =\alpha \, ;  \nonumber\\
W^2 |_{\rho=0} &=&1 \,  ; \, \,  (W^2)^{\prime}|_{\rho=0} =0.
\label{regularity}
\end{eqnarray}
Because of Eqs.(\ref{Lexpansion}) and (\ref{regularity}), a $\delta$-type singularity arises while solving the $(\mu\nu)$ components of the Einstein equations as
\begin{equation}
G_{\mu \nu} |_{\text{\textnormal{sing}}} 
         = - \frac{B^{\,\prime\prime}}{B}\, g_{\mu\nu} = (1- \alpha) \frac{\delta (\rho)  }{ B} \, g_{\mu\nu}.
 \label{deltasource}
\end{equation}
In terms of the $d$-dimensional indexes, this $\delta$-like singular  behavior can be described by an energy-momentum tensor of the form\footnote{Again, we can think about the codimension-one analog of this balance: in that case, in order to capture the effect of the brane as a source, we need to add the Gibbons-Hawking term to the Einstein-Hilbert action, which makes the variational principle well defined \cite{GibbonsHawkingTerm}. }
\begin{equation}\label{T}
	T_{AB} =\delta _{A}^{\mu} \:  \delta _{B}^{\nu} \:   \hat{T}_{\mu\nu}(x)  \frac { \delta (\rho)  } {2 \pi B}.
\end{equation}
Matching the singular contributions (\ref{deltasource}) and (\ref{T}) from both sides of Einstein's equations at the brane position $\rho=0$, we obtain the general form of the $(d-2)$-dimensional density of energy-momentum,
\begin{equation}
\hat{T}_{\mu\nu} (x) = \frac{1}{8\pi\, G_d} \left[2\pi \, (1-\alpha) \right]  \: \hat{g}_{\mu\nu} (x) =  \frac{\Delta \phi}{8\pi\, G_d} \,  \: \hat{g}_{\mu\nu} (x) ,
\label{braneT1}
\end{equation}
where $G_d$ denotes Newton's constant in $d$ dimensions. Thus, we see that the brane must have an energy-momentum tensor proportional to their induced metric \cite{CanadaGroup}. 

Returning to our case, we apply the above prescription  to the image space $\widetilde{dS}_{d}$ through the following changes:
\begin{eqnarray}
\rho(z) &=& 2L \tan^{-1} \left[ \tanh \left(  \frac{z}{2L}    \right)   \right]    \, \, ;  \, \, \,    0 \leq \rho \leq \frac{\pi L}{2} \\
\phi &\rightarrow& \alpha \phi
\label{change}
\end{eqnarray}
obtaining
\begin{equation}
ds^2 = d\rho^2 + \alpha^2 L^2 \sin^2(\rho/L) d\phi^2 + \cos^2(\rho/L)\, \hat{g}_{\mu\nu}(x) dx^{\mu}dx^{\nu} .
\label{newnormal}
\end{equation}
In this coordinates, it is clear that the conditions  (\ref{Lexpansion}) and (\ref{regularity}) are fulfilled, and, consequently, we can consider the $(d-2)$-dimensional submanifold defined by $\hat{g}$ as a gravitational source (a $p$-brane). The geometry of this brane, as mentioned before, should be consistent with Eq.(\ref{braneT1}). Hence, we consider a source with an energy-momentum of the form
\begin{equation}
\hat{T}_{\mu\nu} (x) = \sigma \,\hat{g}_{\mu \nu}(x)
\label{braneT2}
\end{equation}
where $\sigma$ defines brane tension, understood as the energy density contribution of the brane embedding.
Comparing expressions for the energy-momentum tensor (\ref{braneT1}) and (\ref{braneT2}), we obtain the relation between the angular deficit and the internal structure of the brane \cite{FrolovIsraelUnruh}:
\begin{equation}
\sigma  =  \frac{\Delta \phi }{8\pi\, G_d}.
\label{sigma}
\end{equation}

From the brane point of view, Einstein field equations should be fulfilled. Thus, if we denote by $\lambda_{\text{eff}}$  the effective cosmological constant of the brane, then $\hat{g}$ must satisfy $\hat{G}_{\mu\nu} + \lambda_{\text{eff}} \, \hat{g}_{\mu\nu} =0$, where $\lambda_{\text{eff}}$ should takes into account the mean energy density of the empty space (what we usually call cosmological constant $\lambda$), and of the brane tension $\sigma$ given by Eq.(\ref{sigma}). That is
\begin{eqnarray}
\lambda_{\text{eff}} &=& \lambda - 8\pi G_{d-2} \, \sigma .
\label{leff1}
\end{eqnarray}
This can be explicitly calculated by dimensionally reducing the Einstein-Hilbert action using the metric ansatz (\ref{normal}). Starting from 
$d$ dimensions, we get
\begin{eqnarray}
I[g] &=& \frac{1}{16\pi G_d} \int d^d x \sqrt{-g} \, (R-2\Lambda)  \nonumber \\
 &=&\frac{1}{16\pi G_d} \, \int d\rho \, d\theta\, B(\rho) \, W^{d-4}(\rho)  \, \int d^{d-2}x \sqrt{-\hat{g}} \, \left[ \hat{R} - 2 \Lambda \frac{d-4}{d-2} W^2(\rho) \right] \nonumber\\
 &=&\frac{1}{16\pi G_d} \, \left( \frac{2\pi\,\alpha\, L^2}{d-3}\right)  \, \int d^{d-2}x \sqrt{-\hat{g}} \, \left[ \hat{R} - 2 \Lambda \frac{(d-3)(d-4)}{(d-1)(d-2)}\right].
\end{eqnarray}
From the last line, we identify Newton's constant in $(d-2)$ dimensions as
\begin{equation}
G_{d-2} = \left(\frac{d-3}{2\pi \,\alpha \, L^2} \right) G_d.
\label{G}
\end{equation}
Including now the effect of the gravitational source (\ref{braneT2})
\begin{eqnarray}
I[\sigma]= \int d^{d-2} x \sqrt{-\hat{g}} \, \, \sigma
\label{Ib}
\end{eqnarray}
we obtain the effective $(d-2)$-dimensional cosmological constant on the brane
\begin{equation}
\lambda_{\text{eff}} = \frac{(d-3)(d-4)}{(d-1)(d-2)} \, \Lambda  - 8\pi G_{d-2} \,\sigma =: \chi(d) \, \Lambda - 8\pi G_{d-2} \, \sigma .
\label{leff2}
\end{equation}
The latter expression is in agreement with Eq.(\ref{leff1}), providing $\lambda = \chi(d) \, \Lambda$. Moreover, in absence of the source ($\sigma=0$), the $\hat{g}$-submanifold has a positive $\lambda_{\text{eff}}$.
However, for a non vanishing tension, we have three possibles values of it. Combining Eqs.(\ref{sigma}) and (\ref{G}), Eq.(\ref{leff2}) becomes
\begin{equation}
\lambda_{\text{eff}} = \frac{d-3}{2L^2} (d-2-2\alpha^{-1}).
\label{leff3}
\end{equation}
Therefore, the parameter $\alpha$, responsible for regulating the angular deficit, determines the value of the induced cosmological constant,
\begin{equation}
\alpha \,  \left\{ \begin{array}{lr} > \frac{2}{d-2}  & \mbox{dS brane}, \\
 						=\frac{2}{d-2}  & \mbox{(Ricci) flat brane},  \\
                                                        < \frac{2}{d-2}  & \mbox{AdS brane}.
\end{array}\right.
\label{abound}
\end{equation}
This is a useful result because it gives us the freedom to localize, depending on the size of the angular deficit, different types of branes at the origin of the image space. In the next section, we will use this freedom to map the entropy of $dS_d$ into the entropy of the brane black hole.


\section{Entropy}
\label{entropy}
So far, we have constructed a new spacetime $\widetilde{dS_d}$, which is the image of de Sitter under the transformations (\ref{R(z)}), (\ref{brane}) and (\ref{Wick}). The main feature of this geometry is that it can be seen as a black hole localized on the $z=0$ surface. This surface (the brane) is the image of the cosmological horizon  of the original de Sitter space. Having established this correspondence, we conjecture that the entropy of $dS_d$ can be related to the entropy of the localized black hole in the image space.

\subsection{Five dimensional case}
Here, we focus on the case of five dimensions. The induced metric on the brane is described by the following $f$-term:
\begin{equation}
f_{5}(r)= \beta - \mu \pm \frac{r^2}{\ell^2_{\text{eff}}}
\label{f5}
\end{equation}
where the sign of the last term depends on the domain in which $\alpha$ is defined (\ref{abound}).
We will show that the image space $\widetilde{dS}_{5}$ contains a BTZ black hole \cite{BTZ} localized  on a 2-brane \footnote{The construction of a de Sitter 2-brane as a surface of $dS_5$ is possible, too. However, we focus on the study of an anti-de Sitter 2-brane for reasons that will become clear below.}. This brane is in correspondence with the $dS_5$ cosmological horizon, and, therefore we identify the $dS_5$ horizon degrees of freedom with those that result in the BTZ entropy.

In order to carry this procedure out, and according to Eq.(\ref{abound}), we need to impose the $AdS$ bound $\alpha < \frac{2}{3}$. Consequently, the effective cosmological constant will be related with $AdS_3$ effective radius $\ell_{\text{eff}}$ in the usual way: $\lambda_{\text{eff}} = - 1/\ell_{\text{eff}}^2$.
Then, Eq.(\ref{leff3}) gives
\begin{equation}
\ell_{\text{eff}}^2 = \frac{\alpha L^2}{2-3\alpha}.
\label{lvsL}
\end{equation}
Note that for $\alpha=1$, the previous relation is consistent with Eq.(\ref{f}).
Now, in Eq.(\ref{f5}), fixing $\mu$ and $\beta$  in terms of $\ell_{\text{eff}}$ and some characteristic length $r_H$, in the following way:
\begin{equation}
\beta - \mu= -\frac{r_{H}^2}{\ell_{\text{eff}}^2},
\label{betaBTZ}
\end{equation}
the metric for the 2-brane will be
\begin{equation}
d\hat{s}^2=-\left(\frac{r^2 - r_{H}^2}{\ell_{\text{eff}}^2} \right) dt^2 +\left(\frac{r^2 - r_{H}^2}{\ell_{\text{eff}}^2} \right)^{-1} dr^2 + r^2 d\phi^2\label{BTZ}=: ds^2_{BTZ}
\end{equation}
This is the line element for the (nonrotating) BTZ black hole, with event horizon located at $r=r_{H}$. It characterizes the geometry of the 2-brane, acting as a gravitational source embedded within the image space $\widetilde{dS}_{5}$.

Once the embedding is done, transformation (\ref{brane}) can be understood in two steps:  first, it changes the isometry group of the 3-sphere of radius $L$ (the spacelike surface at $R=L$ in $dS_5$), by the $AdS_3$ isometry group. Next, it takes the quotient such that the resulting group submanifold is BTZ black hole\footnote{It is a well-known fact that BTZ black hole is obtained by discretes identifications of $AdS_3$. In terms of the isometry groups, it is given by the quotient $SO(2,2)/ \Sigma$, with $\Sigma$ a discrete soubgroup of $SO(2,2)$. Further details of this construction can be reviewed in Refs.\cite{BHTZ, CarlipTeitelboim}.
}. Schematically, the change $L^2 d\Omega^2_{d-2} \rightarrow d\hat s^2$ defined in Eq.(\ref{brane}), in terms of the respective isometry groups, has the form
\begin{equation}
SO(4) \rightarrow SO(2,2) / \Sigma \cong \text{BTZ}
\label{group}
\end{equation}
where $\Sigma$ denotes a discrete subgroup of SO(2,2)  \cite{BHTZ, CarlipTeitelboim}. Moreover, after applying the previous transformation, and doing the change $\rho=L\theta$ in the metric (\ref{newnormal}), the image space line element takes the form
 \begin{equation}
 ds^2 = L^2(d\theta^2 + \alpha^2 \sin^2 \theta\, d\phi^2) + \cos^2 \theta \, ds^2_{BTZ}
 \end{equation}
 with $0\leq \theta \leq \pi/2$. From here we read off that $\widetilde{dS}_{d}$ has the topology of a (half of a) 2-sphere (up to deficit angle) times a warped BTZ black hole.

Having constructed the image space, let us compute the entropy at  the brane position. Here, the only degrees of freedom that contribute to such entropy, are those from the BTZ event horizon $r_H$. Then, the entropy on the brane will be given by the Bekenstein-Hawking entropy of the BTZ black hole, that is
\begin{equation}
S_{BTZ} = \frac{2\pi\, r_H}{4\,G_3}.
\label{BTZentropy}
\end{equation}
As argued in section \ref{localization}, the only physical effect of fixing the parameter $\alpha$ is the sign of the induced cosmological constant $\lambda_{\text{eff}}$. At this point, there is no loss of generality if we set $\alpha$ to any value consistent with the $AdS$ bound. In particular we can think in\footnote{ Let us analyze the freedom we have for this choice: for the nonrotating version of BTZ black hole the event horizon depends on its mass and of the $AdS_{3}$ radius, in our case, $r_{_{H}}=\sqrt{M}\, \ell_{\text{eff}}$. Consequently, and according with the relation (\ref{lvsL}), setting  $\alpha = r_{H}^{-1} \, L$, is equivalent to pick up one of the roots of the cubic equation $M \alpha^{3} + 3\, \alpha -2 =0 $, such that $\alpha<2/3$. This is always possible.
The previous reasoning makes clear, too, why this choice is not consistent with the localization of a de Sitter 2-brane. If so, we would need, on one hand to take the bound $\alpha > 2/3$, but, on the other hand, in order to have a positive mass $M>0$, we would need $\alpha <0$, which is a contradiction.
}
\begin{equation}
\alpha = r_{H}^{-1} \, L .
\label{alpha5d}
\end{equation}
Replacing  $r_H$ from the above choice into the BTZ entropy (\ref{BTZentropy}), and using Eq.(\ref{G}) to rewrite $G_{3}$ in terms of $G_{5}$, we get\footnote{The $N$ volume of the unitary $N$-sphere  is given by
\begin{equation}
S_N=\frac{2\pi^{\frac{N+1}{2}}}{\Gamma(\frac{N+1}{2})}  \nonumber
\end{equation}
}
\begin{eqnarray}
S_{BTZ} &=& \frac{1}{4\,G_5} \left[ \frac{2\pi^2}{\Gamma(2)}  \right] \, \alpha \, r_{_{H}}L^2
                =\frac{S_3\,L^3}{4\,G_5} = \frac{A_{dS}}{4 \, G_5}
\label{conjecture5d}
\end{eqnarray}
where $A_{dS}$ denotes the area of the $dS_5$ cosmological horizon. The previous equation relates the entropy of the full five-dimensional de Sitter space and that of the BTZ black hole localized on the 2-brane,
\begin{equation}
S_{dS} =  S_{BTZ}.
\label{5d}
\end{equation}
That is, if the brane black hole in the image space is induced through Eq.(\ref{alpha5d}), the entropy of $dS_5$ matchs exactly the Bekenstein-Hawking entropy of the BTZ brane black hole.    

\subsection{Higher dimensional case}
Finally, let us consider the case in which the dimension of the spacetime is higher than or equal to six. Depending on the $\alpha$-bound (\ref{abound}), the brane geometry is that of a Schwarzschild black hole for a null $\lambda_{\text{eff}}$, or a Schwarzschild-(anti-) de Sitter
($S(A)dS_{d-2}$) black hole for a non zero $\lambda_{\text{eff}}$.  However, as in the case of five dimensions, there is a restriction which determines the nature of the brane. For this reason we focus on the embedding of a $SdS_{d-2}$ black hole within the image space. That is, we consider $\alpha > \frac{2}{d-2}$.

At the brane location, the only contribution to the entropy is that coming from the $SdS_{d-2}$ black hole. Denoting by $r_{H}$ the event horizon of the black hole, we have
\begin{equation}
S_{BH}=\frac{A}{4\,G_{d-2}}= \frac{ S_{d-4} \, r_{H} ^{d-4}}{4\,G_{d-2}} .
\label{bhentropy1}
\end{equation}
Using again  the relation between Newton's constants (\ref{G}),  the previous equation takes the form
\begin{equation}
S_{BH} = \left( \frac{2\pi\, \alpha \, L^2}{d-3} \right) \frac{S_{d-4} \, r_{H}^{d-4} }{4\,G_d}.
\label{bhentropy2}
\end{equation}
Now, because the $SdS_{d-2}$ black hole horizon is always located at $r_{_H} < \ell_{\text{eff}}$, necessarily,  $r_{_H} = \gamma \, \ell_{\text{eff}}$, for some $\gamma < 1$.
Therefore, we can take\footnote{This choice is equivalent to fix $\alpha$. Here, the need for a de Sitter brane becomes evident. It follows from the relation between $\ell_{\text{eff}}$ and $L$, that demanding
$\gamma <1$, is equivalent to require $|d-2-2\alpha^{-1}  | < (d-4) \, \alpha^{\frac{2}{d-4}}$. The above inequality is only fulfilled if $\alpha > \frac{2}{d-2}$, which is the bound for a $dS$ brane.}
\begin{equation}
\gamma = \alpha^{\frac{1}{4-d}} \, \left( \frac{L}{\ell_{\text{eff}}} \right) < 1 .
\label{gamma}
\end{equation}
Under this choice, the Bekenstein-Hawking entropy of the brane black hole becomes
\begin{eqnarray}
S_{BH} &=& \left[ \frac{2\pi\, S_{d-4}}{d-3} \right] \frac{L^{d-2}}{4\, G_d}
                         = \left[ \frac{2\pi^{\frac{d-1}{2}}}{ \Gamma(\frac{d-1}{2})}  \right] \,  \frac{L^{d-2}}{4\, G_d}
                        =  \frac{S_{d-2} \, L^{d-2}}{4\, G_d} = \frac{A_{dS}}{4\, G_d}
\label{bhentropy3}
\end{eqnarray}
which is exactly the entropy of the full $dS_d$ spacetime. Then, the entropy of $dS_d$ can be related to the entropy of the localized $SdS_{d-2}$ black hole in the image space,
\begin{equation}
S_{dS} = S_{BH}.
\end{equation}


\section{Concluding Remarks}
\label{CR}
In this paper, we have obtained a relation between the entropy of de Sitter space and the entropy of a lower-dimensional black hole. To reach such relation, we have constructed a map from de Sitter space $dS_d$ into a new spacetime $\widetilde{dS}_{d}$.
Although both spaces have different topologies, they arise from the same action principle. The geometry of the latter describes a brane black hole, localized on a conical defect defined by a deficit angle $\alpha$. 

Under the correspondence $\mathfrak{F}: dS_d \rightarrow \widetilde{dS}_{d}$, the preimage of the brane is precisely the cosmological horizon of de Sitter space  $\mathcal{H}= \mathfrak{F}^{-1}$(brane). Then, due to the holographic principle, the degrees of freedom that originate the $dS_d$ entropy are mapped into the surface defined by the brane.  Within this context it was shown that, for a suitable choice of the deficit angle $\alpha$, the entropy of de Sitter space may be obtained by computing the Bekenstein-Hawking entropy of the brane black hole.

In five dimensions, the $dS_5$ horizon was mapped into an $AdS_3$ slice immerse within the image space $\widetilde{dS_5}$. This slice represents a 2-brane for which the induced metric describes a BTZ black hole.  Using the freedom of fixing the deficit angle consistently with the $AdS$ bound (\ref{abound}), it was shown that $S_{dS} =  S_{BTZ}$.  For dimensions higher than five, in order to match the entropies from both sides of $\mathfrak{F}$, we had to induce a positive cosmological constant on the brane, fixing the value of the parameter $\alpha$ as indicated in Eq.(\ref{gamma}). In this case, the resulting geometry of the image space was that of a Schwarzschild-de Sitter black hole localized on a $(d-3)$-brane. This configuration of the image space led to the relation $S_{dS} = S_{BH}$.

The previous results seem to indicate that the problem of entropy for de Sitter and black hole spaces, different in principle, may be related by a duality. Nevertheless, it will require future work to understand completely the map $\mathfrak{F}$. For instance, one might hope to relate some other observables using this approach. This would provide stronger evidence of the proposed duality, and it represents the most pressing open problem to address in upcoming research.

Along this lines, the further development of the five-dimensional case would be of particular interest. Here, the spatial boundary of $dS_5$ (its horizon, $\mathcal H$) is dual to the $AdS_3$. But, at the same time,
 $AdS_3$ is dual to a two-dimensional conformal field theory \cite{BrownHenneaux, Maldacena}.
Then, one may attempt to use  the sequence  $\mathcal H \subseteq dS_5 \mapsto AdS_3/CFT_2$ in order to understand the entropy of $dS_5$  in terms of the microstates of the $CFT_2$. More ambitiously, it may be explored if the $dS_5$ microstates can be counted in the same way as those of the BTZ black hole \cite{Strominger}, that is, as the asymptotic density of states of the Hilbert space on which the $CFT_2$ is defined, through the Cardy's formula \cite{Cardy}.

Additional directions for future work include, at the quantum level, the investigation of loop corrections to the entropy and how these corrections behave under $\mathfrak{F}$.  At the classical level, the study of the geometrical features of the image space 
$\widetilde{dS_d}$ and the Gregory-Laflamme stability \cite{GL} of the different brane black holes embedded in the image space would be of interest.

It will remain to be further investigated whether the relation between the entropy of de Sitter space with that of a black hole presented here is indeed a geometrical accident. Nevertheless, we consider that this might be a clue that a more robust scenario could be implemented.

\bigskip
\noindent \textit{Acknowledgments:} R.A. likes to thank {\footnotesize  \sc FONDECYT} 1131075 and DI-286-13/R 
({\footnotesize  \sc UNAB}).
N.Z. is partially supported by  {\footnotesize  \sc CONICYT} under the Anillo project {\footnotesize  \sc ACT}1122.



\end{document}